\documentclass[lettersize,journal]{IEEEtran}
\usepackage{amsmath,amsfonts}
\usepackage{algorithmic}
\usepackage{algorithm}
\usepackage{array}
\usepackage[caption=false,font=normalsize,labelfont=sf,textfont=sf]{subfig}
\usepackage{textcomp}
\usepackage{stfloats}
\usepackage{url}
\usepackage{verbatim}
\usepackage{graphicx}
\usepackage{cite}
\usepackage{multirow}
\usepackage[table,xcdraw]{xcolor}

\begin{document}

\title{gpuPairHMM: High-speed Pair-HMM Forward Algorithm for DNA Variant Calling on GPUs}
\author{%
    \IEEEauthorblockN{%
        Bertil Schmidt\IEEEauthorrefmark{1},
        Felix Kallenborn\IEEEauthorrefmark{1},
        Alexander Wichmann\IEEEauthorrefmark{1},
        Alejandro Chacon\IEEEauthorrefmark{2},
        Christian Hundt\IEEEauthorrefmark{2}} \\
    \IEEEauthorblockA{\IEEEauthorrefmark{1}%
            Insitute of Computer Science, Johannes Gutenberg University, Germany
            \{bertil.schmidt,kallenborn,alwichma\}@uni-mainz.de} \\
    \IEEEauthorblockA{\IEEEauthorrefmark{2}%
       NVIDIA Corporation, Santa Clara, USA, 
       \{achacon,chundt}@nvidia.com\}%
       }




\maketitle

\begin{abstract}
The continually increasing volume of DNA sequence data has resulted in a growing demand for fast implementations of core algorithms. Computation of pairwise alignments between candidate haplotypes and sequencing reads using Pair-HMMs is a key component in DNA variant calling tools such as the GATK HaplotypeCaller but can be highly time consuming due to its  quadratic time complexity and the large number of pairs to be aligned. Unfortunately, previous approaches to accelerate this task using the massively parallel processing capabilities of modern GPUs are limited by inefficient memory access schemes. This established the need for significantly faster solutions. We address this need by presenting gpuPairHMM -- a novel GPU-based parallelization scheme for the dynamic-programming based Pair-HMM forward algorithm based on wavefronts and warp-shuffles. It gains efficiency by minimizing both memory accesses and instructions. We show that our approach achieves close-to-peak performance on several generations of modern CUDA-enabled GPUs (Volta, Ampere, Ada, Hopper). It also outperforms prior implementations on GPUs, CPUs, and FPGAs by a factor of at least 8.6, 10.4, and 14.5, respectively. gpuPairHMM is publicly available at \url{https://github.com/asbschmidt/gpuPairHMM}.
\end{abstract}


\section{Introduction}
\IEEEPARstart{D}{uring} 
the last decade we have witnessed a tremendous increase in the volume of data generated in the life sciences, especially propelled by the rapid progress of next generation sequencing (NGS) technologies. It has been estimated that between one hundred million and two billion human genomes could be sequenced by 2025 \cite{stephens2015big} which has a major impact on many areas of life.

Corresponding Illumina sequencing datasets often consist of many short reads of a few hundred base-pairs in length. 
DNA variant calling is an important operation performed on the alignments returned by mapping these sequencing reads to reference genomes in order to identify different types of mutations. Pair Hidden Markov Models (Pair-HMMs) are a key component of variant calling tools including GATK HaplotypeCaller (HC) \cite{gatk_haplo} and MuTect2 \cite{benjamin2019calling}. In a typical use case all pairwise alignments between a set of candidate haplotypes sequences and a set of reads mapped to the considered region are calculated to determine the most likely haplotypes using the Pair-HMM forward algorithm. 
However, the associated time complexity, proportional to the product of alignment target lengths, often makes this algorithm a major contributor to overall runtimes since it has to be executed for a large number of read-haplotype pairs for a typical dataset. Modern genomic sequencing pipelines therefore often require highly efficient implementations. This motivates the need for novel solutions on parallel computing platforms which can keep pace with the increasing volume of genomic sequencing data.

Consequently, a variety of parallelized implementations have been developed for the Pair-HMM forward algorithm in recent years on CPUs \cite{gatk_haplo,GKL,Mauricio,misra2018performance,snytsar2023pairhmm}, GPUs \cite{HMM_GPU_Li2021, branchini2021, ren2018}, FPGAs \cite{rauer2016,huang2017,sampietro2018fpga, wang2019,wertenbroek2019acceleration}, and PIMs \cite{abecassis2023gapim}. However, existing GPU implementations are often limited by inefficient memory access and computation schemes and thus cannot fully exploit the performance of modern GPUs.  Thus, there is an urgent need for a parallelization approach that reaches close-to-peak performance on modern GPUs.

We address this need by presenting gpuPairHMM --- a highly efficient parallelization scheme for accelerating this key algorithm for DNA variant calling from NGS data on CUDA-enabled GPUs. It is based on a wavefront design and employs warp shuffle instructions for efficient inter-thread communication.

In summary, this paper makes the following contributions:
\begin{itemize}
\item Fine-grained parallelization strategy based on warp intrinsics for fast inter-thread communication (instead of using shared or global memory) and minimizing overall executed instructions, (similar techniques have also been applied to other bioinformatics algorithms \cite{schmidt2020cudtw++,achacon2014thread,chacon2014boosting,muller2022anyseq,schmidt2024cudasw++}).
\item A novel scheme to efficiently compute emission probabilities using CUDA shared memory.
\item Advanced input partitioning and sequence length binning strategies to support batches with highly varying sequence lengths occurring in typical real-world databases combined with specialized kernels. 
\item Performance evaluation on four generations of modern GPUs (Volta, Ampere, Ada, Hopper). 
\item We outperform prior Pair-HMM implementations on GPUs, CPUs, and FPGAs by a factor of at least 8.6, 10.4, and 14.5, respectively, using real-world datasets of sequences with variable lengths.
\end{itemize}

The rest of the paper is organized as follow. Section \ref{sec:bg} provides necessary background information about Pair-HMMs and GPUs and reviews related work. Our parallelization scheme is presented in Section \ref{sec:method}. Performance is evaluated in Section \ref{sec:perf}. Finally, Section \ref{sec:conclusion} concludes.

\section{Background}\label{sec:bg}

\subsection{Pair-HMM Forward Algorithm for Variant Calling}
DNA variant calling aims to determine mutations in a sequenced organism from a set of NGS reads. GATK HC first uses local \textit{de novo} assembly to determine a set of candidate haplotypes and corresponding active regions in the reference genome. Each candidate is then further inspected by computing an alignment using the Pair-HMM forward algorithm between the haplotype and each input NGS read that is mapped to the corresponding active reference region.  
The returned likelihood scores are used to determine the candidate haplotypes that are explained by the read data with the highest likelihood.   

Pair-HMMs are frequently used in bioinformatics as a statistical model to study the similarity between two sequences. The Pair-HMM forward algorithm is based on the concept of dynamic programming (DP) and computes the overall alignment probability by summing up the probabilities of all possible pairwise alignments between the two given sequences. The specific Pair-HMM forward algorithm used in GATK HC works as follows.

Consider a read $R = (r_1 r_2 \ldots r_m)$ of length $m$ and a candidate haplotpye $H =  (h_1 h_2 \ldots h_n)$ of length $n$ over the alphabet $\Sigma = \{ A, C, G, T, N\}$.
The Pair-HMM forward algorithm aligns $R$ to $H$ and returns a single score representing the likelihood of $R$ being derived from $H$ using the following recurrence relations for all $1 \le i \le m$, $1 \le j \le n$.   

\begin{align} 
    \label{eq:M}
    \begin{split}
        M(i,j) &= \lambda_{i,j} (\alpha_i M(i-1,j-1) + \beta_i (I(i-1,j-1) \\
               & \qquad + D(i-1,j-1)))
    \end{split}\\
    \label{eq:I}  I(i,j) &= \delta_i M(i-1,j) + \epsilon_i I(i-1,j) \\
    \label{eq:D}  D(i,j) &= \zeta_i M(i,j-1) + \epsilon_i D(i,j-1)
\end{align}

Initialization of the three DP matrices $M, D, I$ is given by $M(i,0) = I(i,0) = D(i,0) = 0$, $M(0,j) = I(0,j) = 0$, and $D(0,j) = 1/n$. 

$\alpha_i$, $\beta_i$, $\delta_i$, $\epsilon_i$, $\zeta_i$ are arrays of transition probabilities depending on the read position $i$. $\lambda_{i,j}$ is the emission probability depending on both read position $i$ and haplotype position $j$ as calculated by  

\begin{align}
\label{eq:lambda}
    \lambda_{i,j} &= 
    \begin{cases}
        1 - Q_i & \text{if } r_i = h_j \\
        Q_i/3 & \text{if } r_i \neq h_j
    \end{cases} 
\end{align}

where $Q_i$ is the base quality score of $R$ at position $i$. 

The overall alignment probability of $R$ and $H$ is then determined by $\sum_{j=1}^n (M_{m,j} + I_{m,j})$.

\begin{figure}[t]
    \centerline{\includegraphics[width=0.4\textwidth]{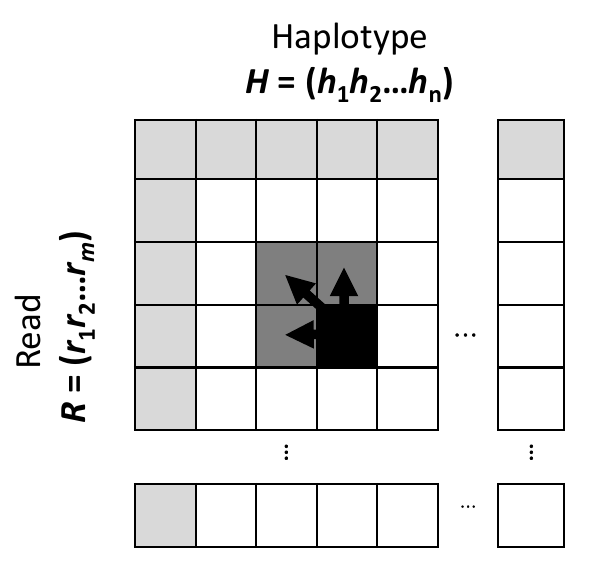}}
    \caption{Conceptual layout of the DP matrix for the Pair-HMM forward algorithm. Light gray cells are initialization cells and dark gray indicates the ancestral subproblems of the currently active cell (black).}
    \label{fig:summary}
\end{figure}

Score-only computations can be performed in linear space $\mathcal{O}(\min\{m, n\})$ and quadratic time $\mathcal{O}(m \cdot n)$.
Note that each cell in the DP matrices $M, D, I$ depends on its left, upper, and upper-left neighbor (see Figure \ref{fig:summary}). 

Input datasets are organized in batches  $I = \{ B_1, \ldots, B_b\}$ where each batch $B_q = (\mathcal{R}_q, \mathcal{H}_q)$  consists of a set of reads $\mathcal{R}_q$ and a set of haplotypes $\mathcal{H}_q$, which need to be compared against each other; i.e. $\rvert \mathcal{R}_q \lvert \times \rvert \mathcal{H}_q \lvert$ alignments per batch. The Pair-HMM forward algorithms thus need to be executed for each batch $q$ resulting in a high computational complexity. 

\subsection{Architectural GPU Features}
CUDA kernels are executed using a number of independent thread blocks. Each thread block is mapped onto exactly one streaming
multiprocessor (SM) and consists of a number of warps. All 32 threads within a warp are executed in lockstep fashion (SIMT). 
CUDA-enabled GPUs contain several types of memory: large but high latency global memory and fast
but small on-chip shared and constant memory. Nevertheless, the fastest way to access data is through the usage of the thread-local register file. 

Modern GPUs provide instructions for warp-level collectives in order to efficiently support communication of data stored in registers between threads within a warp without the need for accessing global or shared memory. A crucial feature of our approach is the usage of warp shuffles for
low latency communication and minimization of memory traffic. In particular, we take advantage of the warp-level collectives 
\texttt{\_\_shfl\_down\_sync()} 
\texttt{\_\_shfl\_up\_sync()}; 
e.g. the intra-warp communication operation 

\texttt{R1 = \_\_shfl\_up\_sync(0xFFFFFFFF,R0,1,32);} 

moves the contents of register \texttt{R0}
in thread $i$ within each warp to register \texttt{R1} in thread $i + 1$ for $0 \le i < 31$.

\subsection{Related Work}

Due to the importance of the Pair-HMM forward algorithm for DNA variant calling it has been optimized on a variety of architectures including CPUs, GPUs, FPGAs, and PIM (Processing in Memory). Even special-purpose customized chips have been recently proposed \cite{gu2023gendp}.

Existing implementations typically feature two levels of parallelism:
\begin{itemize}

\item \textbf{Coarse-grained (Inter-task):} Computation of different alignments in parallel since each alignment within each batch can be calculated independently.

\item \textbf{Fine-grained (Intra-task):} Computation of the
DP matrix of a single alignment in parallel. Note that the dependency relation shown in Figure \ref{fig:summary} (i.e. each cell depends on its left, upper, and upper-left neighbour) allows for cells along a minor diagonal of the DP matrix to be computed in parallel in a wavefront fashion.

\end{itemize}

Intel GKL \cite{GKL, snytsar2023pairhmm, misra2018performance} and PairHMM \cite{Mauricio_workshop} are two CPU-based libraries that feature fine-grained parallelization by taking advantage of AVX2 or AVX-512 vector registers. In addition, Intel GKL features coarse-grained parallelization by means of multi-threading using several CPU cores. 

Ren et al. \cite{ren2018} compared inter-task and intra-task parallelization strategies on GPUs but only achieved a modest speedup of less than 1.2 compared to optimized CPU implementations. Pair-HMM-PP \cite{HMM_GPU_Li2021} was able to improve upon this approach but only report a throughout of up to 69 GCUPS (\underline{G}iga \underline{C}ell \underline{U}pdates \underline{P}er \underline{S}econd) for synthetic datasets of equal length sequences on a V100 GPU. Branchini et al. \cite{branchini2021} improved the performance to 75 GCUPS on a V100 by introducing a dynamic memory swap technique. Nevertheless, all prior GPU-based approaches are limited in performance due to inefficient memory access and computation schemes. In comparison, our approach  presented in this paper can achieve a performance of 590 GCUPS on the same V100 GPU and is able to improve performance on newer GPU generations even further (e.g. up to 2763 GCUPS on an L40S). 

FPGA-based approaches typically exploit fine-grained parallelism by using an array of identical processing elements (PEs) working in systolic fashion. In addition, several of these arrays can be used to exploit coarse-grained parallelism. Corresponding FPGA implementations \cite{sampietro2018fpga, huang2017, wang2019, rauer2016} are able to achieve a performance similar to prior GPU-based implementations.
Furthermore, GAPiM \cite{abecassis2023gapim} is a recent Pair-HMM implementation using the UPMEM-PIM (processing in memory) architecture, that replaces floating-point by lower precision fix-point arithmetic which in turn reduces the accuracy of computed alignment scores. 

In summary, prior acceleration approaches have not been able to provide a significant speedup compared to optimized CPU-based implementations. To address this shortcoming, we thus propose an improved parallelization scheme that can exploit modern GPUs with close-to-peak performance thereby achieving clearly improved speedups.

\section{Parallelization Scheme}\label{sec:method}
\subsection{General Design}

gpuPairHMM is written in C++ and runs on Linux systems.
Using CUDA it can take advantage of GPU accelerators attached to a host CPU (typically via PCIe). More detailed instructions are included in the software repository.

Our algorithm accepts a set $I = \{ B_1, \ldots, B_b\}$ of read-haplotype batches computed by GATK HC as input. Each batch $B_q = (\mathcal{R}_q, \mathcal{H}_q)$  consists of a set of read sequences $\mathcal{R}_q$ and a set of candidate haplotypes $\mathcal{H}_q$, which need to be aligned against each other using the Pair-HMM forward algorithm on the GPU. 
Each read sequence further includes per-base quality scores ($Q_i$), and probabilities for insertion and deletion of gaps as inputs. These values will be used in the kernel to compute the values $\lambda_{i,j},   \alpha_i, \beta_i, \delta_i, \epsilon_i$, and $\zeta_i$. We have further validated that output scores returned by gpuPairHMM are identical to those computed by GATK HC.

Our workflow consists of the following stages:
\begin{enumerate}
    \item Partition $I$ into non-overlapping subsets.
    \item Transfer the first subset of batches from CPU to GPU.
    \item Partition the current subset according to read sequence lengths on the GPU.
    \item Execute optimized GPU alignment kernels for each partition.
    \item Transfer the next subset from CPU to GPU concurrently to alignment kernel computation.
    \item If all subsets have been processed, continue with Stage 7, otherwise continue with Stage 3.
    \item Transfer alignment scores to CPU for output.
\end{enumerate}

Note that our approach allows for concurrent PCIe data transfer and kernel execution by means of CUDA streams. 
Furthermore, it is possible to process datasets exceeding the GPU memory capacity by processing only subsets of $I$.

Our design is based on computing an independent read-haplotype alignment per (sub)warp -- a group of synchronized threads executed in lockstep that can efficiently communicate by means of warp shuffles.
Threads in a (sub)warp compute DP matrix cell values in cooperative fashion. Note that according to Eqs \ref{eq:M}, \ref{eq:I}, \ref{eq:D} each DP cell depends on its left, upper, and upper-left neighbour (see Figure \ref{fig:summary}), which means that (parallel) computation of DP cells has to follow this topological order.

Our hybrid coarse/fine-grained parallel design proposed in this work combined with optimizations described in the following sections leads to important advantages compared to prior GPU-based approaches such as avoiding thread divergence, dynamic partitioning of large memory footprints on local memories, and reduction of the overall executed instructions.
More specifically, we apply the following techniques:
\begin{itemize}
    \item Full in-register computation of recurrence relations.
    \item Low-latency communication of neighboring DP cells between threads through warp shuffles.
    \item Efficient schemes for computing emission and transition probabilities of the PairHMM.
    \item Reduction of frequent sequence character loading from memory by employing an intra-warp communication scheme based on warp shuffles.
\end{itemize}

\subsection{GPU-based Alignment Kernel}

\begin{figure*}[!tpb]
\centerline{\includegraphics[width=0.9\textwidth]{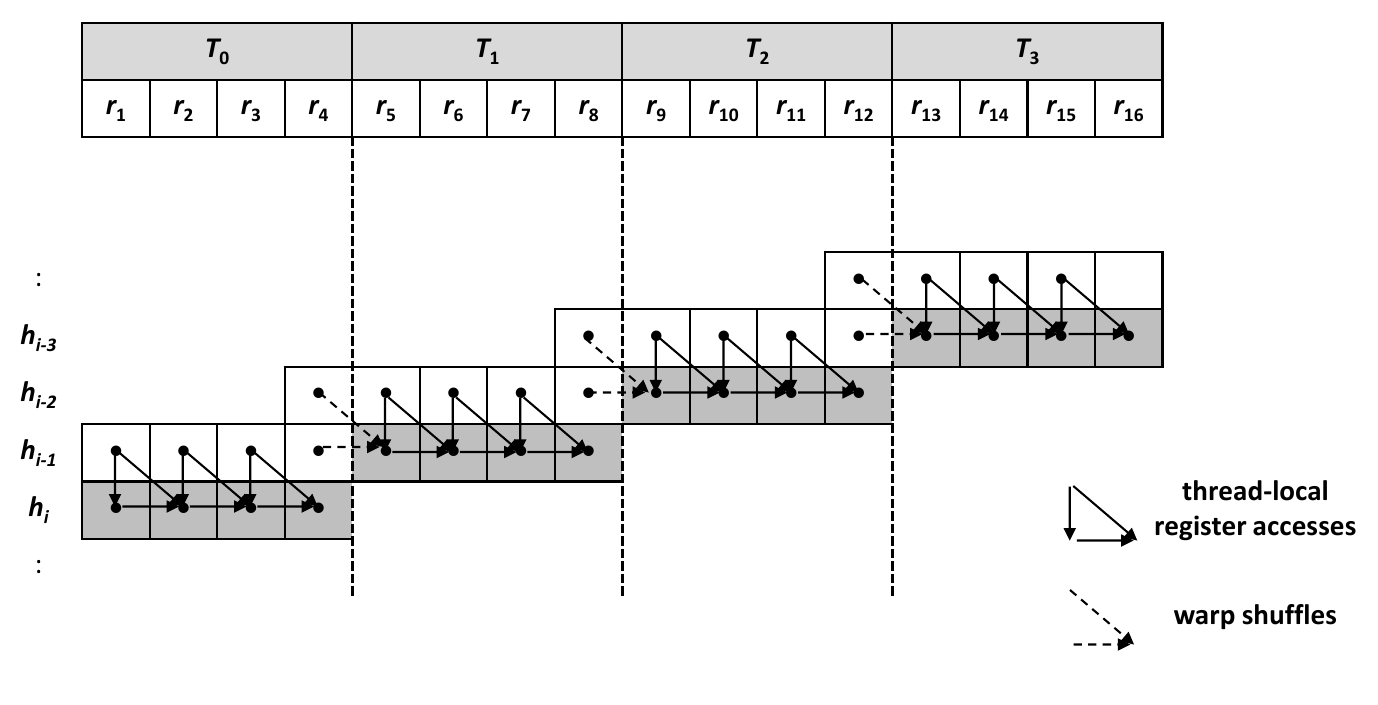}}
\caption{Example of mapping DP matrix computation for comparing $R=(r_1 r_2 \ldots r_m)$ length of $m = 16$ and $H =(h_1 h_2 \ldots h_n)$ of length $n=k \cdot p$ to a group of $p=4$ threads using $k=4$.
Each thread computes $k$ matrix cells (gray) in each iteration $i$ using the values of the antecedent row (white) and an additional value from the second antecedent row.
Threads communicate values in the rightmost of their $k$ columns (as well as the required characters of $H$) by warp shuffle operations (dashed arrows).}
\label{fig:mapping}
\end{figure*}

Our alignment kernels are executed with a large number of
(sub)warps: groups of $p$ threads $T_0,...,T_{p-1}$ with $p \in \{2, 4, 8, 16, 32\}$ are
executed in lockstep. Each such sub(warp) computes the Pair-HMM forward algorithm of a read $R=(r_1 r_2 \ldots r_m)$ of length $m =  k \cdot p$ and a candidate haplotype sequence $H =(h_1 h_2 \ldots h_n)$ of length $n$. Note that, if the length of a read is shorter than $k \cdot p$ it needs to be filled up with 'N' characters.
We assign $k$ columns of the DP matrix to be calculated to each thread.
Computation proceeds along a {\it wavefront} in $m+p$ iterations.
In iteration $i$, thread $T_t$ computes $k$ adjacent cells of the DP matrix row $i-t$.
All cells of the current and previous iteration are stored in thread-local registers. The required access of thread $T_t$ to the rightmost value of thread
$T_{t-1}$ computed in the previous iteration is accomplished by using
the low-latency warp shuffle instruction \texttt{\_\_shfl\_up\_sync()}. Our mapping strategy is illustrated in Figure \ref{fig:mapping}.

At the beginning each thread $T_t, 0 \leq t < p$ loads $k$ characters of $R$ (i.e. $r_{t \cdot k+1} \ldots r_{t \cdot (k + 1)}$) from GPU global memory.
Note that the loaded characters of $R$ remain the same for each column of the DP matrix while the required characters from $H$ vary during computation.

The corresponding per-base quality scores and deletion/insertion probabilities are loaded by each thread in a similar way. Those are then used to compute the emission and transition probabilities. First, $\alpha_i, \beta_i, \delta_i, \epsilon_i$, and $\zeta_i$ are calculated for all $t \cdot k+1\le i \le t \cdot (k + 1)$ in each thread $T_t$ and stored in local registers. Calculation of $\lambda_{i,j}$ is, however, more complex since it depends on  $r_i, Q_i$, and $h_j$. Computing it for each DP cell would involve expensive operations including an evaluation of a conditional statement, which is inefficient on SIMD architectures like GPUs. 

A key insight (which has not been used by prior parallelization approaches) is that there are only five possible values for $h_j$; i.e. $h_j \in \{A, C, G, T, N\} = \Sigma$ for all $1 \le j \le n$. We use this fact to pre-compute all possible values of $\lambda_{i,j}$ in a 2-dimensional array $E$ of size $\lvert \Sigma \rvert \times n$, where $E[c][i]$ stores the value for $\lambda_{i,j}$ with $h_j = c$. Note that computing all values in $E$ only has a complexity of  $\mathcal{O}(\vert \Sigma \rvert \cdot n)$ instead of $\mathcal{O}(m \cdot n)$. We thus compute $E$ by all threads in subwarp in parallel and store it in CUDA shared memory. During the DP matrix computation the required values for $\lambda_{i,j}$ can then be retrieved by looking up the value $E[h_j][i]$ in CUDA shared memory. Furthermore, the required values by each thread for $t \cdot k+1\le i \le t \cdot (k + 1)$ are stored consecutively, which leads to high shared memory access speed. As a consequence, the time for memory lookup can be completely hidden by computation and thus comes at zero cost.


At the start of each iteration, thread $T_0$ of each subwarp loads the required haplotype value $h_j$ to the register $c_{h\_\textrm{current}}$.
After the iteration is completed, a warp shuffle-up of $c_{h\_\textrm{current}}$ is performed to shift the required haplotype letters to neighboring threads.
Note that by reading \texttt{char4} values we are able to further reduce the frequency of loading values of $H$ from global memory by a factor of 4.

To compute $k$ DP matrix cells per iteration, each thread looks up the required emission scores for $\lambda$ from CUDA shared memory and subsequently computes the recurrence relations (Eqs. \ref{eq:M}, \ref{eq:I}, \ref{eq:D}).
This involves several multiplication and  addition (single-precision) floating-point operations using only thread-local in-register computation. Those are further optimized using three more efficient multiply-add instructions.   
Subsequently, the rightmost DP cell computed in Thread $t$ is communicated to Thread $t+1$ using the low latency warp shuffle instruction \texttt{\_\_shfl\_up\_sync()}. Neighboring query sequence letters are also communicated between threads in the same way (see Figure \ref{fig:mapping}). 

To complete iteration $k$ each thread continuously updates the values $\sum_{j=1}^k (M_{m,j} + I_{m,j})$ for each of its assigned columns $m$ in an accumulator registers. At the end of each haplotype-read alignment computation the accumulator register is written to CUDA global memory as output.

Using appropriate values of $k$ and $p$ in our algorithm is critical to optimize execution times for specific ranges of read sequence length. 
We have thus declared $k$ and $p$ as C++ template parameters which allows us to generate various GPU kernels tailored to different read lengths ranges and (sub)warp sizes. For example, using $k=32$ and $p=8$ ($k=32$ and $p=4$) would generate a kernel that aligns read sequences of length up to 256 (128) with the appropriate number of registers to maximize performance for that combination of sequence length and GPU. 

\subsection{Input partitioning}

\begin{figure}[htbp]
\centerline{\includegraphics[width=0.5\textwidth]{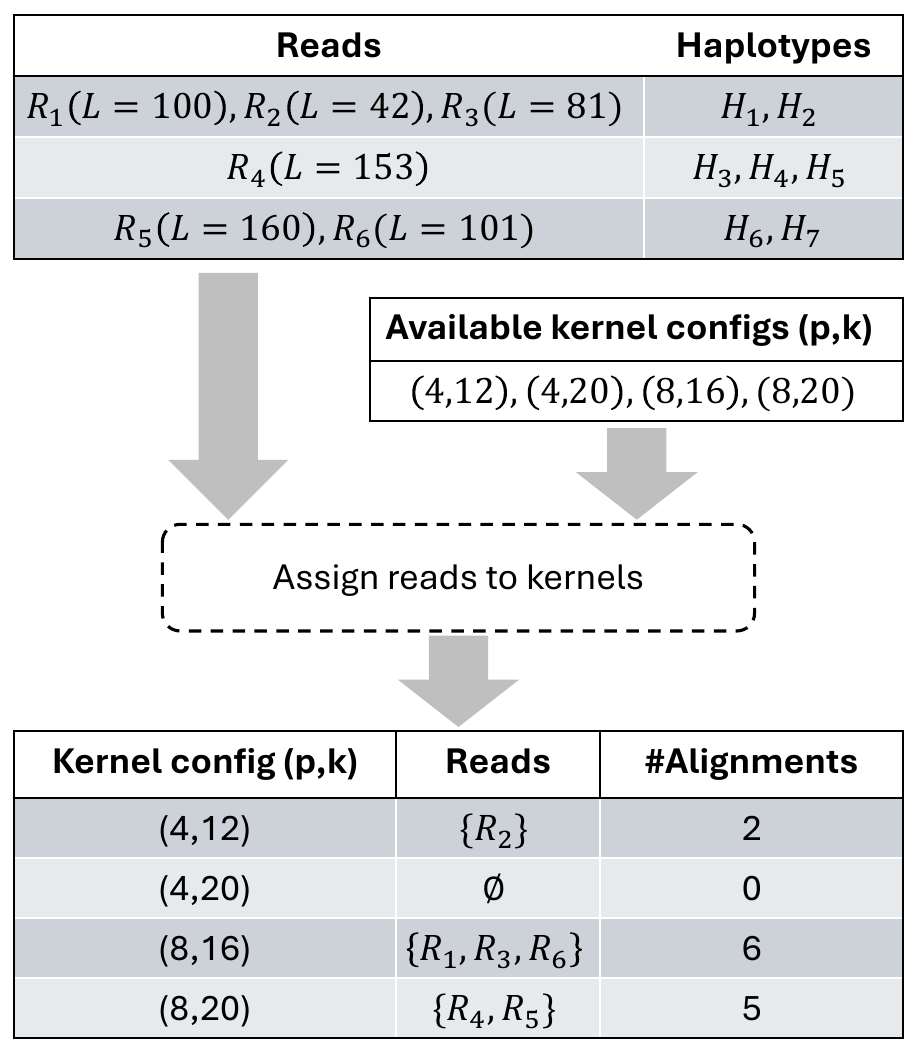}}
\caption{Example of partitioning a list of three batches $ B_1 = (\{ R_1, R_2, R_3\},\{H_1, H_2 \}), B_2 = (\{ R_4 \}, \{ H_3, H_4, H_5 \}), B_3 = (\{ R_5, R_6 \}, \{ H_6, H_7 \})$ depending on the read lengths given using set of four available kernel configurations. The maximum read length for each kernel is $M=p \cdot k$. For each read, the smallest possible $M$ is selected. After partitioning, the corresponding kernels will be launched using one sub-warp per alignment. }
\label{fig:partitioning}
\end{figure}

Recall that the input to our algorithm is a list of batches where each batch $B_i$ consists of a set $\mathcal{R}_i$ of reads and a set $\mathcal{H}_i$ of haplotypes that should be aligned to each other.
For best performance each read sequence should be processed by a kernel with suitable $k$ and $p$ configuration parameters.
Thus, as a preprocessing step before executing the Pair-HMM algorithm, we determine the list of sequences to be processed for each set of supported kernel parameters. To be more specific, we create a separate index list for each kernel referencing the corresponding sequences in the input data. This is illustrated in Figure \ref{fig:partitioning}. 
Then, for each non-empty index list the corresponding kernel is launched passing the index list as argument to indicate which reads to process.
The individual kernels are executed in distinct CUDA streams which enables concurrent kernel execution in cases where index lists are short.

A typical partitioning by sequence length could use a dedicated kernel for each combination $(p,k) \in \{4,8,16, 32\} \times \{4,8,12,16,20,24,28,32\}$. This would enable fast processing of sequences up to a length of 1,024 bps, which is sufficient for typical datasets.

\section{Results}\label{sec:perf}

\subsection{Experimental Design}

We compared gpuPairHMM to different publicly available CPU- and GPU-based Pair-HMM implementations GATK \cite{gatk_haplo} (CPU), Intel GKL \cite{GKL} (CPU), PairHMM \cite{Mauricio} (CPU), and Pair-HMM-PP (GPU) \cite{HMM_GPU_Li2021}. 
For Intel GKL two versions were tested, the GATK-integrated version and a standalone variant, which is the base GKL version with a custom reader bypassing the Java code. 

Furthermore, we compared performance to a number of other approaches that are not publicly available by taking performance numbers from the respective papers for GPUs \cite{HMM_GPU_Li2021, branchini2021} and FPGAs \cite{rauer2016,huang2017,sampietro2018fpga, wang2019}.

We evaluated the performance on the following systems:
\begin{description}

\item[\textbf{S1:}] Dual-socket Intel Xeon Gold 6238 CPU ($2\times$22 physical cores, 88 logical cores) $2.1$GHz, capable of AVX2 and AVX-512 instructions and 188GB of RAM with a V100 GPU attached.

\item[\textbf{S2:}] AMD Threadripper 3990X CPU (64 physical cores, 128 logical cores) $2.90$GHz, capable of AVX2 instructions, 256GB of RAM and a L40S GPU (350 watts TDP).

\item[\textbf{S3:}] AMD EPYC 7713P CPU (64 physical cores, 128 logical cores) $2.00$GHz with 512GB RAM and attached with an A100 GPU (80GB).

\item[\textbf{S4:}] Intel Xeon Silver 4314 CPU @ 2.40GHz with 256GB of RAM and attached with an NVIDIA H100 GPU (700 watts TDP).

\item[\textbf{S5:}] AMD EPYC 7313P 16-core CPU with a low-power L4 GPU (72 Watts TDP).

\end{description}

nvcc~12.6 and gcc~9.3.0 are used as GPU and host compiler, respectively, on all systems.

We used three types of datasets.
\begin{itemize}
    \item {\bf Synthetic:} To study the peak performance that can be achieved by our kernels, we have generated synthetic datasets consisting of large batches of reads and haplotpyes of identical length.

    \item {\bf 10s:} "10s" is a well-known (small) dataset available from \cite{Mauricio} that has been used to measure performance of many prior approaches. It consists of $3,550$ read-haplotype pairs divided into $7$ batches with read length varying from $10$ to $247$. The average read length is $55$.

    \item {\bf NA12878:} Large-scale dataset generated by GATK HC by performing variant calling for a subset of the NA12878 sample of the $1,000$ genomes project \cite{1000genome}. 
    It consists of $55,125,211$ read-haplotype pairs divided into $998,837$ batches with read length varying between $10$ and $151$ and haplotype length between $30$ and $521$. The average read length is $58$.
\end{itemize}

\subsection{Kernel Performance}
We first evaluate the performance of our CUDA kernels on four recent GPU generations (Volta, Ampere, Hopper, Ada) in an ideal situation where all read sequences are of same length and ideally fit the corresponding kernel template parameters. To identify the maximum achievable performance and best performing template parameters value for $(p,k)$ we just measure kernel execution times; i.e. we exclude any overheads like PCIe data transfers and we do not consider varying  sequence lengths. 

Speeds for Pair-HMM tools are often reported by measuring runtimes and converting them into the number of DP matrix cell updates that are performed per second; i.e., {\bf TCUPS} (\underline{T}rillions of \underline{C}ell \underline{U}pdates \underline{P}er \underline{S}econd) as:
\begin{equation}
\mathrm{TCUPS} = \frac{\sum_i m_i \times n_i} {t \times 10^{12}}  \label{eq:peakTCUPS}
\end{equation}
where $t$ is the runtime in seconds and $m_i$ and $n_i$ are the lengths of the alignment targets.

\begin{figure}[t]
\centerline{\includegraphics[width=0.52\textwidth]{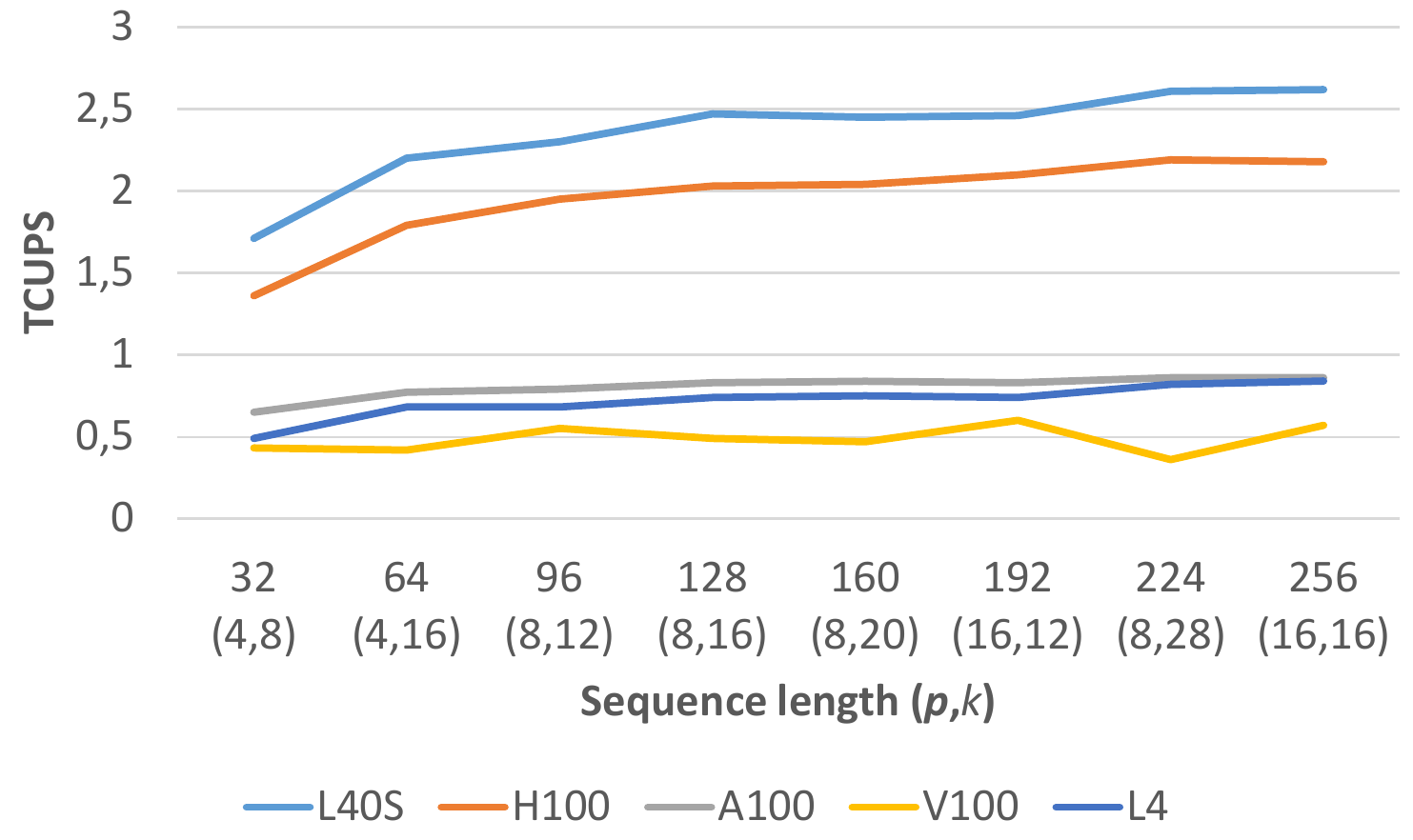}}
\caption{Peak performance (in TCUPS) of our alignment kernels scanning synthetic databases with all sequences of identical length on  different GPUs (L40S, H100, A100, V100, L4). The corresponding template parameter configurations are also indicated $(p,k) =  (\textrm{thread group size},\textrm{number of columns})$}
\label{fig:Peak}
\end{figure}

Figure \ref{fig:Peak} shows the achieved peak performance in terms of TCUPS for varying sequence lengths with the corresponding best performing template parameters for each length.
It can be seen that performance generally increases with sequence length (with some variation for V100 due to register pressure in some cases) with peak performance of 2.62 TCUPS, 2.37 TCUPS, 0.86 TCUPS, 0.59 TCUPS, and 0.84 TCUPS for L40S, H100, A100, V100, and L4, respectively. This result also shows that gpuPairHMM can benefit from increased computational power of new-generation GPUs.

\subsection{Performance Comparison to other Tools}

Table \ref{tab:NA} compares the performance of gpuPairHMM to other publicly available tools on GPUs (Pair-HMM-PP) and CPUs (GATK, PairHMM, Intel GKL) for the NA12878 dataset. 
Measured GPU runtimes now include PCIe data transfer times in addition to various other overheads (length partitioning, sorting). GATK is executed with both its original sequential JAVA implementation and with logless caching optimization. The sequential CPU-based PairHMM is executed using AVX2 vectorization on S1. Furthermore, the highly optimized Intel GKL CPU-based approach is executed using AVX2 vectorization with 128 threads on S2 and AVX-512 vectorization using 88 threads on S1. gpuPairHMM can achieve a speedup of 44 over the best performing CPU-based implementations (Intel GKL, 128 threads) and of 174 compared to GPU-based Pair-HMM-PP. 

Despite highly varying sequence lengths in the NA12878 dataset (read length is ranging between 10 and 151 with an average of 58) and the various other overheads (data transfers, partitioning, sorting), gpuPairHMM achieves a runtime of only 0.69 seconds which gives a performance of 1.82 TCUPS on an L40S. This corresponds to 83\% of the overall peak kernel performance for all equal sequences of length 64 (the kernel configuration closest to the average length of this dataset) without any overheads. This demonstrates the effectiveness of our overall workflow (see Figure \ref{fig:partitioning}).

\begin{table}[!t]
\caption{Runtimes (in seconds) of different CPU and GPU implementations for the NA12878 dataset including PCIe data transfers for GPU implementations. All GPU implementations were executed on S2 (with L40S GPU). CPU implementations were tested on S1 for AVX512 and on S2 otherwise. Speedups of gpuPairHMM compared to other implementations and the number of utilized CPU threads are also indicated. \label{tab:NA}}
\centering
\begin{tabular}{|l||l||c||r|}
\hline
Tool  & Hardware & Runtime &  gpuPairHMM \\
  &  &  {[}s{]}  &  Speedup\\
\hline
{\bf gpuPairHMM} & & {\bf 0.69} & 1.0 \\
Pair-HMM-PP   & \multirow{-2}{*}{GPU (S2)} & 120  & 174\\
\hline
GATK (Original)  & CPU (S2, 1) & 65,387 & 94,764 \\
GATK (Caching) & CPU (S2, 1) & 6,017 & 8,720 \\
PairHMM & CPU (S2, AVX2, 1) & 2,008  & 2,910 \\
Intel GKL & CPU (S2, AVX2, 128) & 30 & 44 \\
Intel GKL &  CPU (S1, AVX512, 88)  & 37  & 54 \\
\hline
\end{tabular}
\end{table}

Table \ref{tab:10s} shows the runtimes of various tools on CPUs, GPUs, and FPGAs for the well-known 10s dataset. Since most of the tested implementations are not publicly available, corresponding runtimes were taken from the respective literature indicated as references. gpuPairHMM was executed on L40S, H100, A100, V100, and L4 again including PCIe data transfers. It can be seen that gpuPairHMM is able to outperform all previous approaches. When using the same hardware (V100), we are able to outperform the second best GPU-based implementation \cite{branchini2021} by a factor of 8.6. Furthermore, on a V100, we outperform the best FPGA and CPU implementation by a factor of 14.5 and 10.4, respectively. In addition, the performance of gpuPairHMM can be even further increased by a factor of 1.6, or 3.9 when using a newer A100 or L40S, respectively.

\begin{table}[!t]
\caption{Runtimes of different tool using the 10s dataset on different CPUs, FPGAs, and GPUs. The runtimes are taken form the respective paper, when a reference is given.\label{tab:10s}}
\centering
\begin{tabular}{|c||c||r||r|}
\hline
                        & Design {[}Reference{]}            & Runtime    & Speedup to\\
                        &                                   & {[}ms{]}   & Java baseline\\ 
\hline
                        & Java baseline \cite{gatk_haplo}   & 10.800     & 1\\
                        & GKL (S2, AVX2, 128)               & 1.580      & 6,837\\
\multirow{-3}{*}{CPU}   & GKL (S1, AVX512, 88)              & 1.740      & 6,206\\
   
\hline                
                        & Intel Stratix V \cite{rauer2016}  & 8.300      & 1,301\\
                        & Intel Stratix V \cite{huang2017}  & 5.300      & 2,038\\
                        & AMD Xilinx KU3 \cite{sampietro2018fpga}   & 5.000      & 2,160\\
                        & Intel Arria 10 \cite{rauer2016}   & 2.800      & 3,857\\
                        & Intel Arria 10 \cite{huang2017}   & 2.600      & 4,154\\
\multirow{-6}{*}{FPGA}  & Intel Arria 10 \cite{wang2019}    & 2.200      & 4,909\\
\hline                
                        & K40  \cite{ren2018}               & 24.900     & 434\\
                        & V100 \cite{HMM_GPU_Li2021}        & 13.800     & 783\\
\multirow{-3}{*}{GPU}   & V100 \cite{branchini2021}         & 1.300      & 8,154\\

\hline        
              & {\bf V100}   & {\bf 0.152} & {\bf 71,053} \\
    {\bf New: GPU }            & {\bf L4}   & {\bf 0.118} & {\bf 91,525} \\
 {\bf gpuPairHMM }      & {\bf A100}     & {\bf 0.096} & {\bf 112,500} \\        
   & {\bf  H100}   & {\bf 0.050}  & {\bf 216,000} \\    
              & {\bf  L40S} & {\bf 0.039}  & {\bf  276,923} \\               
    

\hline
\end{tabular}
\end{table}


\section{Conclusion}\label{sec:conclusion}

The continually increasing volume of sequence data has resulted in a growing demand for fast implementations of core algorithms such as those for DNA variant calling. 
Computation of the probability that each candidate haplotype can be explained by the observed
data using the Pair-HMM forward algorithm is a key component of the frequently used GATK HC.
However, due to its quadratic time complexity per alignment it is a major contributor to overall runtime.

In this paper, we have presented gpuPairHMM, a fast implementation of the Pair-HMM forward algorithm, which gains high speed on modern GPUs architectures based on warp shuffles for communicating data between threads, and exploiting the latest GPU features. Furthermore, we incorporated a novel way to compute emission probabilities effectively by taking advantage of fast CUDA shared memory and designing a workflow that can efficiently handle PCIe data transfers and real-world datatsets with variable sequence lengths. 

As a consequence, our implementation achieves a theoretical peak performance of 2.62 TCUPS for long sequences of all equal length and a performance of 1.82 TCUPS for real-world datasets including overheads. We outperform previous GPU-based approaches by one-to-two orders-of-magnitude. 
Furthermore, gpuPairHMM offers a clear advantage over state-of-the-art CPU-based approaches.
Using a single L40S GPU it outperforms the highly optimized Intel GKL implementation running on a dual-socket 44-core Intel workstation with 88 threads and a 64-core AMD workstation with 128 threads by 54$\times$ and 44$\times$, respectively. Moreover, all prior reported FPGA-accelerated implementations are outperformed by at least a factor of 14.5. 
We evaluated performance on a variety of different GPUs. In terms of runtime performance L40S performed best. However, low-power GPUs like the L4 might also be an attractive option when considering energy-efficiency. Considering the TDP of 72W for an L4 and 350W for an L40S yields to a performance of up to  11.9 GCUPS/Watt for an L4 and 7.8 GCUPS/Watt for an L40S.

Our results also show that prior GPU approaches have not been able to unlock the full potential of GPUs for the Pair-HMM forward algorithm and only achieve a fraction of the available floating-point performance. As a consequence, they cannot provide significant speedups compared to current CPU-based approaches since they are bottlenecked by inefficient memory access schemes. Our approach thus changes the standing of GPUs for DNA variant calling significantly. We demonstrate that the Pair-HMM algorithm can be re-designed to be fully compute bound, reaching high performance on modern GPUs architectures, with a peak performance of 2.62 TCUPS. 

Our parallelization scheme may also be applicable for accelerating other DP-based bioinformatics algorithms on GPUs with similar dependency relationships such as the Viterbi algorithm for profile hidden Markov models \cite{anderson2024fpga} or computation of multiple sequence alignments based on pairwise alignments \cite{gonzalez2016msaprobs}. It would thus be interesting to evaluate the performance of our approach when adapted to different DP algorithms. 

gpuPairHMM is publicly available as open-source software at \url{https://github.com/asbschmidt/gpuPairHMM}.



%
\bibliographystyle{IEEEtran}
\bibliography{references}

\begin{thebibliography}{10}
\providecommand{\url}[1]{#1}
\csname url@samestyle\endcsname
\providecommand{\newblock}{\relax}
\providecommand{\bibinfo}[2]{#2}
\providecommand{\BIBentrySTDinterwordspacing}{\spaceskip=0pt\relax}
\providecommand{\BIBentryALTinterwordstretchfactor}{4}
\providecommand{\BIBentryALTinterwordspacing}{\spaceskip=\fontdimen2\font plus
\BIBentryALTinterwordstretchfactor\fontdimen3\font minus \fontdimen4\font\relax}
\providecommand{\BIBforeignlanguage}[2]{{%
\expandafter\ifx\csname l@#1\endcsname\relax
\typeout{** WARNING: IEEEtran.bst: No hyphenation pattern has been}%
\typeout{** loaded for the language `#1'. Using the pattern for}%
\typeout{** the default language instead.}%
\else
\language=\csname l@#1\endcsname
\fi
#2}}
\providecommand{\BIBdecl}{\relax}
\BIBdecl

\bibitem{stephens2015big}
Z.~D. Stephens, S.~Y. Lee, F.~Faghri, R.~H. Campbell, C.~Zhai, M.~J. Efron, R.~Iyer, M.~C. Schatz, S.~Sinha, and G.~E. Robinson, ``Big data: astronomical or genomical?'' \emph{PLoS biology}, vol.~13, no.~7, p. e1002195, 2015.

\bibitem{gatk_haplo}
\BIBentryALTinterwordspacing
R.~Poplin, V.~Ruano-Rubio, M.~A. DePristo, T.~J. Fennell, M.~O. Carneiro, G.~A. Van~der Auwera, D.~E. Kling, L.~D. Gauthier, A.~Levy-Moonshine, D.~Roazen, K.~Shakir, J.~Thibault, S.~Chandran, C.~Whelan, M.~Lek, S.~Gabriel, M.~J. Daly, B.~Neale, D.~G. MacArthur, and E.~Banks, ``Scaling accurate genetic variant discovery to tens of thousands of samples,'' \emph{bioRxiv}, 2018. [Online]. Available: \url{https://www.biorxiv.org/content/early/2018/07/24/201178}
\BIBentrySTDinterwordspacing

\bibitem{benjamin2019calling}
D.~Benjamin, T.~Sato, K.~Cibulskis, G.~Getz, C.~Stewart, and L.~Lichtenstein, ``Calling somatic snvs and indels with mutect2,'' \emph{BioRxiv}, p. 861054, 2019.

\bibitem{GKL}
P.~Foley, A.~Prabhakaran, K.~Gururaj, M.~Naik, S.~Gopalan, A.~Shargorodskiy, and E.~Brau, ``Accelerate genomics research with the broad-intel genomics stack,'' 2017.

\bibitem{Mauricio}
M.~Carneiro, ``Optimization of a haplotype pairhmm class for gpu/fpga and avx processing,'' https://github.com/MauricioCarneiro/PairHMM, 2013.

\bibitem{misra2018performance}
S.~Misra, T.~C. Pan, K.~Mahadik, G.~Powley, P.~N. Vaidya, M.~Vasimuddin, and S.~Aluru, ``Performance extraction and suitability analysis of multi-and many-core architectures for next generation sequencing secondary analysis,'' in \emph{Proceedings of the 27th International Conference on Parallel Architectures and Compilation Techniques}, 2018, pp. 1--14.

\bibitem{snytsar2023pairhmm}
R.~Snytsar, ``Pairhmm improvements for modern instruction set architectures,'' in \emph{2023 IEEE International Conference on Bioinformatics and Biomedicine (BIBM)}.\hskip 1em plus 0.5em minus 0.4em\relax IEEE, 2023, pp. 3328--3331.

\bibitem{HMM_GPU_Li2021}
E.~Li, S.~S. Banerjee, S.~Huang, R.~K. Iyer, and D.~Chen, ``Improved gpu implementations of the pair-hmm forward algorithm for dna sequence alignment,'' in \emph{2021 IEEE 39th International Conference on Computer Design (ICCD)}, 2021, pp. 299--306.

\bibitem{branchini2021}
B.~Branchini, A.~Zeni, and M.~D. Santambrogio, ``A methodology for accelerating variant calling on gpu,'' in \emph{SC23: The International Conference for High Performance Computing, Networking, Storage, and Analysis}, 2021.

\bibitem{ren2018}
S.~Ren, K.~Bertels, and Z.~Al-Ars, ``Efficient acceleration of the pair-hmms forward algorithm for gatk haplotypecaller on graphics processing units,'' \emph{Evolutionary Bioinformatics}, vol.~14, p. 1176934318760543, 2018.

\bibitem{rauer2016}
C.~Rauer and N.~Finamore, ``Accelerating genomics research with opencl and fpgas,'' \emph{Altera, Now Part of Intel, Tech. Rep}, 2016.

\bibitem{huang2017}
S.~Huang, G.~J. Manikandan, A.~Ramachandran, K.~Rupnow, W.-m.~W. Hwu, and D.~Chen, ``Hardware acceleration of the pair-hmm algorithm for dna variant calling,'' in \emph{Proceedings of the 2017 ACM/SIGDA International Symposium on Field-Programmable Gate Arrays}, 2017, pp. 275--284.

\bibitem{sampietro2018fpga}
D.~Sampietro, C.~Crippa, L.~Di~Tucci, E.~Del~Sozzo, and M.~D. Santambrogio, ``Fpga-based pairhmm forward algorithm for dna variant calling,'' in \emph{2018 IEEE 29th International Conference on Application-specific Systems, Architectures and Processors (ASAP)}.\hskip 1em plus 0.5em minus 0.4em\relax IEEE, 2018, pp. 1--8.

\bibitem{wang2019}
P.~Wang, Y.~Lei, and Y.~Dou, ``Pair-hmm accelerator based on non-cooperative structure,'' \emph{IEICE Electronics Express}, vol.~16, no.~15, pp. 20\,190\,402--20\,190\,402, 2019.

\bibitem{wertenbroek2019acceleration}
R.~Wertenbroek and Y.~Thoma, ``Acceleration of the pair-hmm forward algorithm on fpga with cloud integration for gatk,'' in \emph{2019 IEEE International Conference on Bioinformatics and Biomedicine (BIBM)}.\hskip 1em plus 0.5em minus 0.4em\relax IEEE, 2019, pp. 534--541.

\bibitem{abecassis2023gapim}
N.~Abecassis, J.~G{\'o}mez-Luna, O.~Mutlu, R.~Ginosar, A.~Moisson-Franckhauser, and L.~Yavits, ``Gapim: Discovering genetic variations on a real processing-in-memory system,'' \emph{bioRxiv}, pp. 2023--07, 2023.

\bibitem{schmidt2020cudtw++}
B.~Schmidt and C.~Hundt, ``cudtw++: ultra-fast dynamic time warping on cuda-enabled gpus,'' in \emph{Euro-Par 2020: Parallel Processing: 26th International Conference on Parallel and Distributed Computing, Warsaw, Poland, August 24--28, 2020, Proceedings 26}.\hskip 1em plus 0.5em minus 0.4em\relax Springer, 2020, pp. 597--612.

\bibitem{achacon2014thread}
A.~Chac{\'o}n, S.~Marco-Sola, A.~Espinosa, P.~Ribeca, and J.~C. Moure, ``Thread-cooperative, bit-parallel computation of levenshtein distance on gpu,'' \emph{Proceedings of the 28th ACM international conference on Supercomputing}, pp. 103--112, 2014.

\bibitem{chacon2014boosting}
A.~Chac{\'o}n, S.~Marco-Sola, A.~Espinosa, P.~Ribeca, and J.~Moure, ``Boosting the fm-index on the gpu: Effective techniques to mitigate random memory access,'' in \emph{IEEE/ACM transactions on computational biology and bioinformatics}, 2014, pp. 1048--1059.

\bibitem{muller2022anyseq}
A.~M{\"u}ller, B.~Schmidt, R.~Membarth, R.~Lei{\ss}a, and S.~Hack, ``Anyseq/gpu: a novel approach for faster sequence alignment on gpus,'' in \emph{Proceedings of the 36th ACM International Conference on Supercomputing}, 2022, pp. 1--11.

\bibitem{schmidt2024cudasw++}
B.~Schmidt, F.~Kallenborn, A.~Chacon, and C.~Hundt, ``Cudasw++ 4.0: ultra-fast gpu-based smith--waterman protein sequence database search,'' \emph{BMC bioinformatics}, vol.~25, no.~1, p. 342, 2024.

\bibitem{gu2023gendp}
Y.~Gu, A.~Subramaniyan, T.~Dunn, A.~Khadem, K.-Y. Chen, S.~Paul, M.~Vasimuddin, S.~Misra, D.~Blaauw, S.~Narayanasamy \emph{et~al.}, ``Gendp: A framework of dynamic programming acceleration for genome sequencing analysis,'' in \emph{Proceedings of the 50th Annual International Symposium on Computer Architecture}, 2023, pp. 1--15.

\bibitem{Mauricio_workshop}
M.~Carneiro, ``Acceleration variant calling,'' in \emph{Intel Genomic Sequencing Pipeline Workshop}, 2013.

\bibitem{1000genome}
{1000 Genomes Project Consortium} \emph{et~al.}, ``A global reference for human genetic variation,'' \emph{Nature}, vol. 526, no. 7571, p.~68, 2015.

\bibitem{anderson2024fpga}
T.~Anderson and T.~J. Wheeler, ``An fpga-based hardware accelerator supporting sensitive sequence homology filtering with profile hidden markov models,'' \emph{BMC bioinformatics}, vol.~25, no.~1, p. 247, 2024.

\bibitem{gonzalez2016msaprobs}
J.~Gonz{\'a}lez-Dom{\'\i}nguez, Y.~Liu, J.~Touri{\~n}o, and B.~Schmidt, ``Msaprobs-mpi: parallel multiple sequence aligner for distributed-memory systems,'' \emph{Bioinformatics}, vol.~32, no.~24, pp. 3826--3828, 2016.

\end{thebibliography}

\begin{IEEEbiography}[{\includegraphics[width=1in, height=1.25in, clip, keepaspectratio]{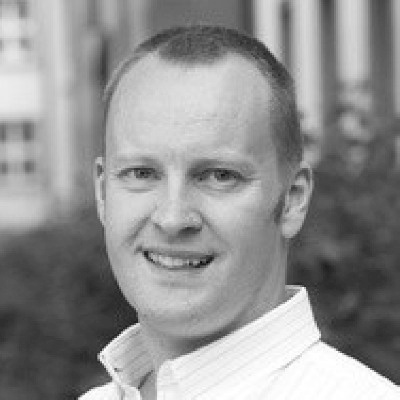}}]{
Bertil Schmidt (Senior Member, IEEE) is currently a tenured full professor and the chair of parallel and distributed architectures with the University of Mainz, Germany. Prior to that, he was a faculty member with Nanyang Technological University, Singapore, and with the University of New South Wales. His research group has designed a variety of algorithms and tools for bioinformatics, mainly focusing on the analysis of large-scale sequence and short read datasets, and data mining. For his research work, he was the recipient of CUDA Research Center Award, CUDA Academic Partnership Award, CUDA Professor Partnership Award, and Best Paper Awards at IEEE ASAP 2009, IEEE ASAP 2015, and IEEE HiPC 2020.}
\end{IEEEbiography}

\begin{IEEEbiography}[{\includegraphics[width=1in, height=1.25in, clip, keepaspectratio]{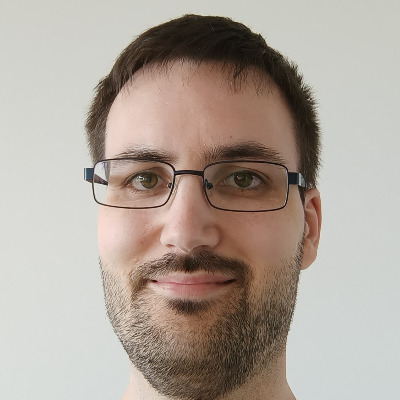}}]{
Felix Kallenborn obtained his PhD in computer science at the Johannes-Gutenberg University in Mainz.
In his current position as post-doctoral researcher at the institute of computer science at JGU Mainz, 
he focuses on the development of CUDA-enabled parallel algorithms in the field of bioinformatics.}
\end{IEEEbiography}

\begin{IEEEbiography}[{\includegraphics[width=1in, height=1.25in, clip, keepaspectratio]{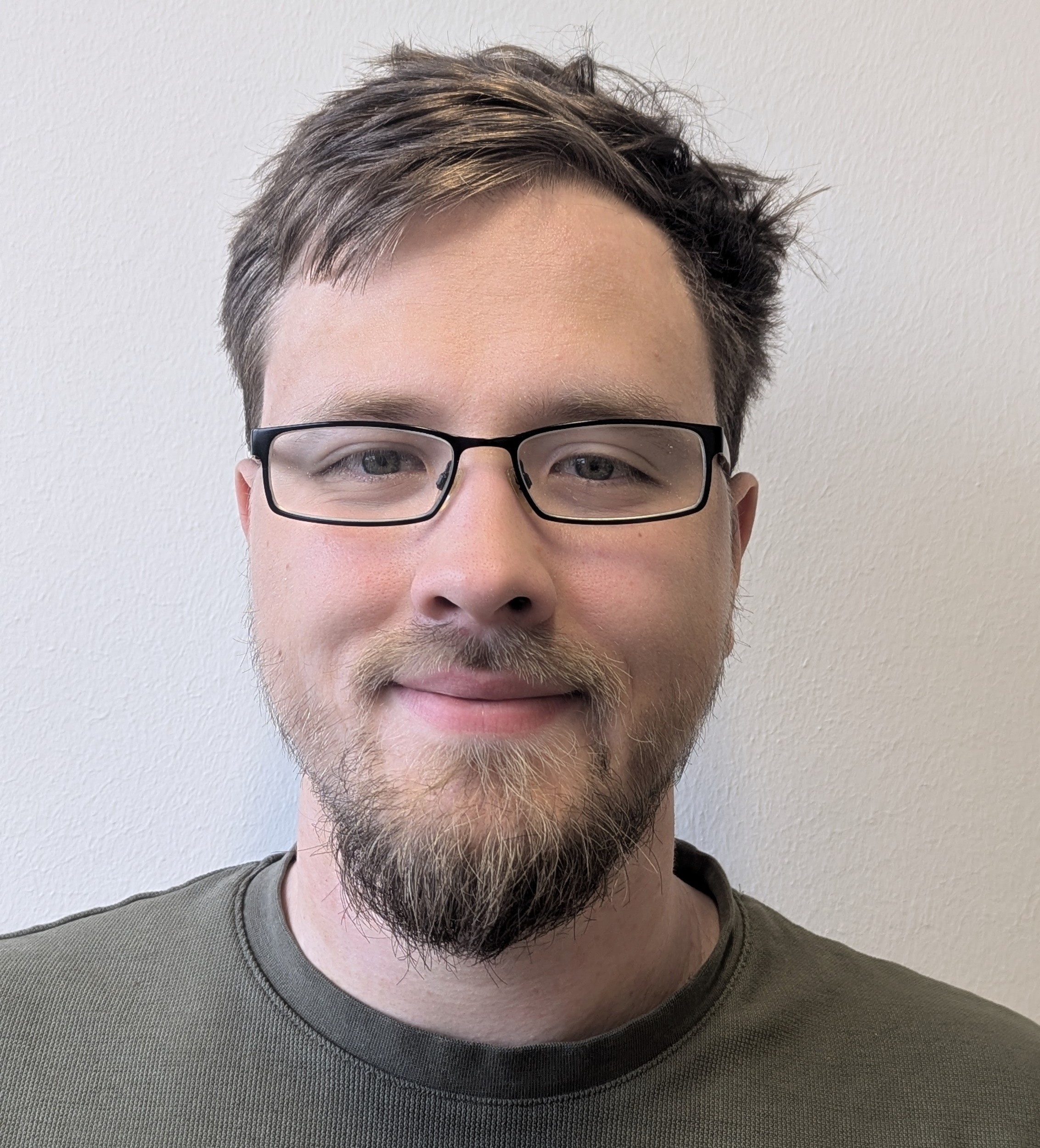}}]{
Alexander Wichmann obtained his master's degree in computer science with a minor in biology. He is currently a PhD student in the HPC group at the institute of computer science at the Johannes Gutenberg University in Mainz, Germany. His work in bioinformatics and chemoinformatics with a focus on Deep Learning.}
\end{IEEEbiography}

\begin{IEEEbiography}[{\includegraphics[width=1in, height=1.25in, clip, keepaspectratio]{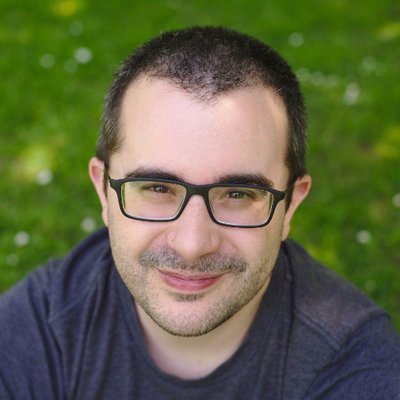}}]{
Alejandro Chacon received his PhD in Computer Science at the Autonomous University of Barcelona, contribuiting to bioinformatic algorithms accelerated on GPUs. He is working on Healthcare at NVIDIA by leading Genomics efforts as Developer Technologies engineer. He is focused in future NVIDIA architectures and researching new algorithms with special interest on Computer Architecture and parallel optimizations for heterogeneous HPC systems. Previously, he was responsible of multiple bioinformatics projects at ARM and Xilinx that lead to product solutions. He developed GEM-cutter, which implements efficient GPU building block primitives used in sequencing production centers.}
\end{IEEEbiography}

\begin{IEEEbiography}[{\includegraphics[width=1in, height=1.25in, clip, keepaspectratio]{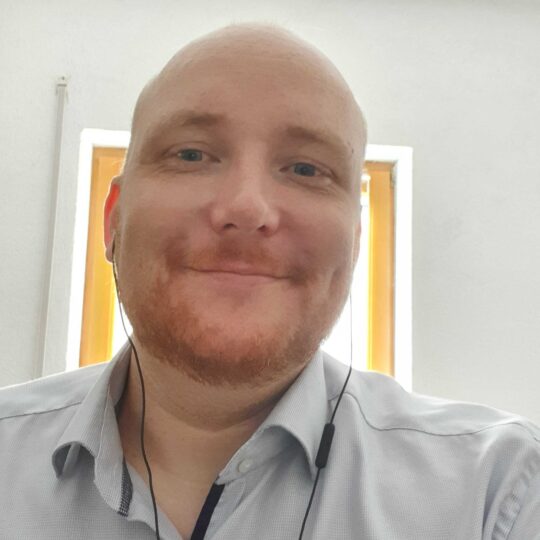}}]{
Christian Hundt received his master's degree in theoretical Physics and his PhD in Computer Science both at the Johannes Gutenberg University in Mainz. In his current role at NVIDIA, heading the Deep Learning Developer Technology EMEAI organization, he leads several teams for (i) efficient multi-modal inference of generative models on CUDA-enabled accelerators, (ii) large-scale training of Deep Learning workloads and performance projections to next-gen GPUs, and (iii) a team for hybrid HPC+AI workloads augmenting traditional algorithmic design with Deep Learning surrogates.  }
\end{IEEEbiography}

\vfill

\end{document}